\documentclass[aps,twocolumn,prd,superscriptaddress,nofootinbib]{revtex4-2}

\usepackage{mathtools}
\usepackage{amsfonts}
\usepackage{mathrsfs}

\usepackage{bbm}
\usepackage{slashed}
\usepackage{amsmath}
\usepackage{bm}
\usepackage{tensor}

\usepackage{graphicx}
\usepackage{color}
\usepackage{colortbl}
\usepackage{array}
\usepackage[dvipsnames]{xcolor}

\usepackage{float}
\usepackage{placeins}
\usepackage{booktabs}
\usepackage{subcaption}
\usepackage{makecell}
\usepackage{tabstackengine}

\usepackage{xspace}
\usepackage{siunitx}
\usepackage{hyperref}
\usepackage[nameinlink]{cleveref}
\usepackage{bookmark}

\usepackage{xifthen}
\usepackage{xcolor}
\usepackage{enumitem}
\hypersetup{
	colorlinks,
	linkcolor={red!75!black},
	citecolor={blue!75!black},
	urlcolor={blue!75!black}
}

\usepackage[utf8]{inputenc}

\setkeys{Gin}{width=0.48\textwidth}
\makeatletter
\newsavebox\myboxA
\newsavebox\myboxB
\newlength\mylenA

\newcommand*\xoverline[2][0.75]{%
	\sbox{\myboxA}{$\m@th#2$}%
	\setbox\myboxB\null
	\ht\myboxB=\ht\myboxA%
	\dp\myboxB=\dp\myboxA%
	\wd\myboxB=#1\wd\myboxA
	\sbox\myboxB{$\m@th\overline{\copy\myboxB}$}
	\setlength\mylenA{\the\wd\myboxA}
	\addtolength\mylenA{-\the\wd\myboxB}%
	\ifdim\wd\myboxB<\wd\myboxA%
		\rlap{\hskip 0.5\mylenA\usebox\myboxB}{\usebox\myboxA}%
	\else
		\hskip -0.5\mylenA\rlap{\usebox\myboxA}{\hskip 0.5\mylenA\usebox\myboxB}%
	\fi}
\makeatother

\graphicspath{{}}

\newcommand{\gettitle}{The causal structure of the quark propagator}

\newcommand{\getHeidelbergAffiliation}{\affiliation{Institut f\"ur Theoretische Physik, Universit\"at Heidelberg, Philosophenweg 16, 69120 Heidelberg, Germany}}

\newcommand{\getEMMIAffiliation}{\affiliation{ExtreMe Matter Institute EMMI, GSI, Planckstr. 1, 64291 Darmstadt, Germany}}

\hypersetup{
	colorlinks,
	linkcolor={red!75!black},
	citecolor={blue!75!black},
	urlcolor={blue!75!black},
	pdftitle={\gettitle},
	pdfauthor={Pawlowski, Wessely},
	pdfkeywords={analytic continuation}
		{correlations functions} {QCD}
		{functional equations}
		{real time} {spectral function}
	bookmarksopen=true,
	bookmarksopenlevel=2,
	bookmarksnumbered=true
}

\allowdisplaybreaks
\begin{document}
\title{\gettitle}

\author{Jan M. Pawlowski}
\getHeidelbergAffiliation
\getEMMIAffiliation

\author{Jonas Wessely}
\getHeidelbergAffiliation

\begin{abstract}

	We study the causal structure of the quark propagator with the spectral DSE. The spectral gap equation is solved with the input of the spectral representation of the gluon  and a causal STI-construction for the quark-gluon vertex. The latter includes a potential infrared enhancement of the vertex strength of the classical tensor structure that accommodates for the physical strength of chiral symmetry breaking. We find a critical vertex strength, below which the quark has a Källén-Lehmann representation. While the nature of the first singularity does not change above the critical strength, we find that the quark propagator features at least two additional pairs of complex conjugate poles that are located approximately at the sum of quark pole mass and peak position of the quark-gluon coupling. These additional poles lead to violations of causality, if they persist in $S$-matrix elements. 
    While the vertex strength of the classical tensor structure in full QCD is below the critical one, that of commonly used vertex models, which rely solely on the classical vertex structure, is typically above it. Finally, we discuss how these additional poles could be avoided in full QCD, where part of chiral symmetry breaking is generated by the other tensor structures in the quark-gluon vertex. 

\end{abstract}

\maketitle

\section{Introduction}
\label{sec:introduction}

The study of dynamical processes in QCD, such as bound-state formation, transport properties of the quark-gluon plasma, or the calculation of form factors and scattering amplitudes, requires access to fully non-perturbative real-time correlation functions. A direct computation of these quantities is only possible within functional approaches such as Dyson-Schwinger equations (DSE) and the functional renormalisation group. The diagrammatic relations in these approaches are governed by the propagators of the elementary and emergent degrees of freedom. Moreover, in bound state computations, e.g.~with Bethe-Salpeter equations (BSE) or Faddeev equations, the propagators, and specifically their causal structure, play a pivotal rôle and have to be used self-consistently. For a respective review, see \cite{Eichmann:2016yit}.

In the present work, we discuss the causal structure of the quark propagator with a focus on its qualitative change with the interaction strength of the classical tensor structure in the quark-gluon vertex. In particular, we will show that beyond a critical interaction strength, the quark propagator features an additional pair of complex conjugated (cc) poles, qualitatively changing its causal structure. The vertex strength of the full QCD vertex is safely below the critical one. Moreover, subject to a spectral representation of the combination of the gluon propagator and quark-gluon vertex, the quark propagator has no cc poles at all. In contrast, that of commonly used vertex models typically lies above the critical one. These additional cc poles are a stable property in the latter approaches and are directly related to the over-enhancement and shape of the Euclidean dressing of the classical tensor structure that defines the quark-gluon avatar of the strong coupling. In conclusion, the over-enhancement of the vertex that is required in the absence of further tensor structure, see e.g.~\cite{Mitter:2014wpa, Williams:2014iea, Williams:2015cvx, Cyrol:2017ewj, Gao:2021wun}, may trigger acausal structures in the system. In turn, in computations with the full quark-gluon vertex, the vertex strength of the classical tensor structure is below the critical one, and part of the full strength of spontaneous chiral symmetry breaking is stored in the non-classical tensor structures. It is suggestive that this combination may avoid the additional cc poles and the related causality problems. 

Our analysis is done within the spectral Dyson-Schwinger equation (DSE) \cite{Horak:2020eng} for the inverse quark propagator. The gap equation for the quark propagator is solved with the input of a spectral gluon propagator \cite{Horak:2021syv, Horak:2023xfb} and a quark-gluon vertex derived from its STI. In particular, this vertex does not introduce unphysical poles or cuts in the complex plane. It is modelled such, that it reproduces the peak position of a physical quark-gluon coupling in the infrared, and we track qualitative changes in the causal structure of the quark propagator depending on the strength of the vertex, i.e., the height of the peak.

In \Cref{sec:quarkgapequ}, we introduce the spectral Dyson-Schwinger equation for the inverse quark propagator and discuss the causal structure for the input correlation functions. We then discuss the emerging complex structure of the solution of the gap equation in \Cref{sec:Results}. It is evaluated  under which conditions the quark propagator has a Källén-Lehmann representation and how additional complex conjugate poles emerge when we enhance the strength of the quark-gluon vertex. Finally, we argue that our results suggest the existence of a  Källén-Lehmann representation of the quark propagator in full QCD and discuss the impact on phenomenological applications in \Cref{sec:conclusion}.

\section{The spectral quark gap equation}
\label{sec:quarkgapequ}

The framework of spectral Dyson-Schwinger equations allows for a self-consistent numerical computation of the quark self-energy in the full complex frequency plane. It has been developed in \cite{Horak:2020eng} and put to use in \cite{Horak:2022myj, Horak:2021pfr, Horak:2022aza} for Yang-Mills theories. For further applications and extensions of the framework to the functional renormalisation group (fRG) see  \cite{Fehre:2021eob, Braun:2022mgx, Horak:2023hkp, Eichmann:2023tjk}. Complementary approaches and earlier real-time computations in a variety of fields with the fRG can be found in  \cite{Gasenzer:2007za, Gasenzer:2010rq, Floerchinger:2011sc, Kamikado:2013sia, Tripolt:2013jra, Pawlowski:2015mia, Kamikado:2016chk, Jung:2016yxl, Pawlowski:2017gxj, Wang:2017vis, Tripolt:2018jre, Tripolt:2018qvi, Corell:2019jxh, Huelsmann:2020xcy, Jung:2021ipc, Tan:2021zid, Heller:2021wan, Roth:2021nrd, Roth:2023wbp}. For further real-time computations in QCD or QED utilising (generalised) spectral representations see  e.g.~\cite{Delbourgo:1977jc, Jia:2017niz, Sauli:2018gos, Solis:2019fzm, Mezrag:2020iuo}, for related work using contour deformations see e.g.~\cite{Stainsby:1990fh,Maris:1991cb,Windisch:2016iud}. 

\begin{figure}[t]
    \centering
    \begin{minipage}{0.99\linewidth}
        \includegraphics[width=\textwidth]{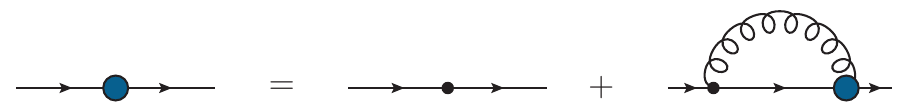}
        \caption{Dyson-Schwinger equation for the inverse quark propagator. Big blue blobs denote full n-point functions, while classical n-point functions are represented by small dots with n legs attached. Full propagators are denoted by straight (quarks) and spiralling (gluons) lines.\hspace*{\fill}  }
        \label{diag:gapequation}
    \end{minipage}
\end{figure}

The full inverse quark propagator is uniquely parametrised by
\begin{align}
    \Gamma^{(q \bar{q})}(p)= Z_q(p)\left[\textrm{i} \slashed p   +  M_q(p) \right]\,,
    \label{eq:inversequarkprop}
\end{align}
with the quark wave function $Z_q(p)$ and the quark mass function $M_q(p)$. The subscript ${}_q$ of the dressing functions refers to the light and strange quark flavors $_l$ and $_s$ respectively. The (inverse) propagator is diagonal in flavor space, and we focus on the light quarks with $(2+1)$ flavor input in this work. We use Euclidean conventions for gamma matrices and momenta,
i.e.~$\lbrace \gamma_\mu, \gamma_\nu \rbrace = 2 \delta_{\mu \nu}$ and $p^2 = p_0^2 + \vec p\,^2$. Real frequencies will be denoted by $\omega$ to distinguish them from their Euclidean counterparts. The inverse propagator satisfies a one-loop exact Dyson-Schwinger equation
\begin{align}
    \Gamma^{(q \bar{q})}(p) & = \textrm{i} Z_2 \slashed p  + Z_m m_q + \Sigma_q(p).
\label{eq:quarkDSE}
\end{align}
The self energy $\Sigma_q(p)$ in \labelcref{eq:quarkDSE} comprises the quantum fluctuations of the quark propagator, for a diagrammatic representation, see \Cref{diag:gapequation}. The renormalisation constants $Z_2$ and $Z_m$ are the wave function and mass renormalisation constants that appear in the renormalised classical action. The self-energy has a diagrammatic representation in terms of full correlation functions and the classical quark-gluon vertex $\textrm{i} g_s \gamma_\mu\mathbbm{1}$ with the gauge coupling $g_s$, and $\mathbbm{1}$ is the unity in flavor and color space. Then, the unrenormalised self energy takes the rather simple form
\begin{align}
    \Sigma_q(p)= g_s C_f Z_1 (\textrm{i} \gamma_\mu) \int_q G_A^{\mu \nu}(q)\, G_q(p+q)\Gamma_\nu(q,p)\,, 
    \label{eq:selfenergy}
\end{align}
where we have omitted  flavor and color indices as the gap equation is color and flavor diagonal and the loop integral is abbreviated as $\int_q = \int \frac{d^4 q}{(2\pi)^4}$. \Cref{eq:selfenergy} depends on the full quark and gluon propagators $G_q(p+q)$ and $G_A^{\mu \nu}(q)$, as well as the full quark-gluon vertex $\Gamma_\nu(q,p)$. The Casimir of the fundamental representation $C_f[SU(3)]=4/3$ and the vertex renormalisation constant $Z_1$ enters the gap equation as a prefactor.

%

\subsection{Spectral structure of quark and gluon propagators}
\label{sec:SpecQuarkGluon}

We solve the Dyson-Schwinger equation in the full complex frequency plane by utilising spectral representations of the full correlation functions involved. This allows for an analytic computation of the momentum integrals of a given loop including the renormalisation, and reduces the computation to a finite spectral one.

The quark propagator $G_q(p)$ has a scalar and Dirac part with
\begin{align}
  G_q(p) = -\textrm{i} \slashed p \, G_q^{d}(p) + G_q^{s}(p)= \frac{-\textrm{i} \slashed p  + M_q(p)}{Z_q(p)\bigl[p^2  + M_q^2(p)\bigr]} \,.
    \label{eq:quarkprop}\end{align}
Both parts have a generalised Källén-Lehmann representation that also accommodates possible additional complex conjugate poles and cuts, induced by  non-analyticities away from the real frequency axis. In the absence of the latter, the spectral representation of the quark propagator takes a very 
compact form \cite{Delbourgo:1977jc}
\begin{align}
    G_q(p) = \int_{-\infty}^{\infty} \frac{d \lambda}{ 2\pi} \frac{\rho_q(\lambda)}{\textrm{i} \slashed p + \lambda}\,,
    \label{eq:quarkspecvac}
\end{align}
where the spectral functions of the scalar and Dirac part of the quark propagator $\rho_q^{(s/d)}$ are given by the antisymmetric and symmetric parts of the quark spectral function $\rho_q$, respectively.
They are defined via the imaginary part of the retarded propagator on the real frequency axis as
\begin{align}\nonumber
    \rho_q^{(d)}(\omega) & = 2\omega \,\text{Im}\, G_q^{d}(p_0\to - \textrm{i} \omega_+)\,, \\[2ex]
    \rho_q^{(s)}(\omega) & = 2 \,\text{Im} \,G_q^{s}(p_0\to -\textrm{i} \omega_+)\,,
    \label{eq:specfuncs}\end{align}
with real frequencies $\omega$, and $\omega_+= \omega + \textrm{i} 0^+$ denotes the retarded limit.

In general, the spectral functions feature delta functions or peaks with a given width and a continuous part that decays for large spectral values. The latter part arises from branch point singularities and continuous cuts and relates to a continuum of scattering states. The former part originates from isolated poles, related to (quasi) particle excitations with or without a decay width. In the following, we choose the Ansatz of a single pole and a scattering continuum for the quark spectral functions,
\begin{align}
    \rho^{(d/s)}_{q}(\lambda) = \pi R_q^{(d/s)} \delta\left(\lambda - m_{q,\text{pole}}\right) + \tilde{\rho}_q^{(d/s)}(\lambda)\, ,
\label{eq:pole-continuum-split}
\end{align}
which anticipates the following analytic structure of the quark self energy: firstly, a zero crossing of the real part of the inverse propagator at the pole mass $m_{q,\text{pole}}$. Secondly a branch cut starting at  $\lambda_\text{scat} \geq m_{q,\text{pole}}$, with the additional constraint that the imaginary part of the mass function approaches zero for $ \omega  \to m_{q,\text{pole}}$. Note, that although the emergence of complex conjugate poles for certain parameter values violates the spectral representation~\labelcref{eq:quarkspecvac}, the spectral functions as defined in~\labelcref{eq:specfuncs} from the cuts on the real axis, does not deviate from the pole-scattering split \labelcref{eq:pole-continuum-split}.
The spectral representation \labelcref{eq:quarkspecvac} entails the sum rules
\begin{align}
    \int_\lambda \rho_{q}^{(d)}(\lambda) & =\lim_{p^2\to \infty} \frac{1}{Z_q(p)}\,,  \qquad  \int_\lambda \lambda \, \rho_{q}^{(s)}(\lambda) = 0\,,
    \label{eq:sum rules}
\end{align}
with
\begin{align}
    \int_\lambda = \int_0^\infty \frac{d\lambda}{\pi}\,.
    \label{eq:SpecInt}
\end{align}
The sum rules \labelcref{eq:sum rules} can be deduced from the asymptotic behaviour of the propagator, for details see \cite{Horak:2022myj}. Note, that the mass function $M_q(p)$ decays logarithmically for large momenta for finite current quark masses, while it decays polynomially with $1/p^2$ in the chiral limit. In both cases, the scalar sum rule in \labelcref{eq:sum rules} holds true.
\begin{figure*}[ht]
    \centering
    \begin{subfigure}[b]{.48\linewidth}
        \centering
        \includegraphics[width=.99\textwidth]{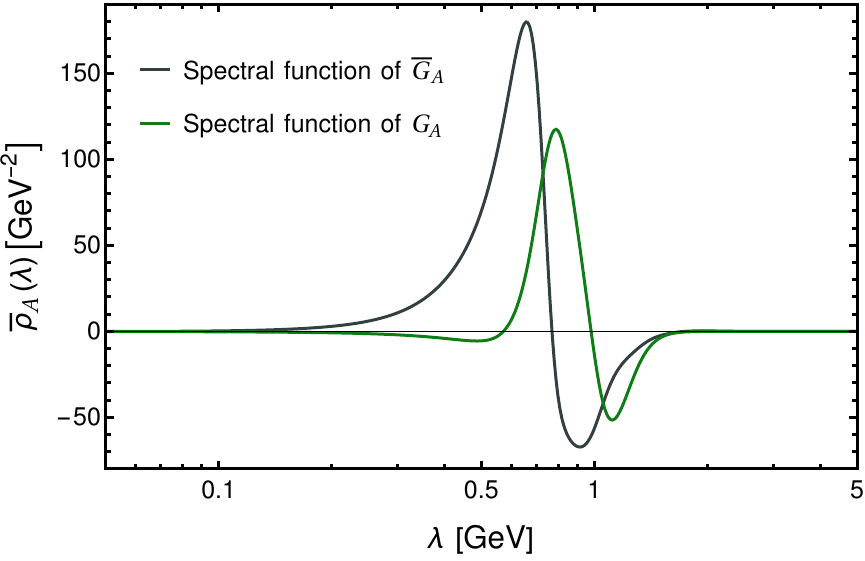}
        \caption{Spectral function of $\bar G (p) = G_A(p)/Z_c(p)$ and $G_A(p)$ respectively. The ghost dressing shifts the peaked structures of the gluon spectral function to the infrared.\hspace*{\fill}}
        \label{fig:gluondressing}
    \end{subfigure}
    \hspace*{.31cm}
    \begin{subfigure}[b]{.48\linewidth}
        \centering
        \includegraphics[width=.97\textwidth]{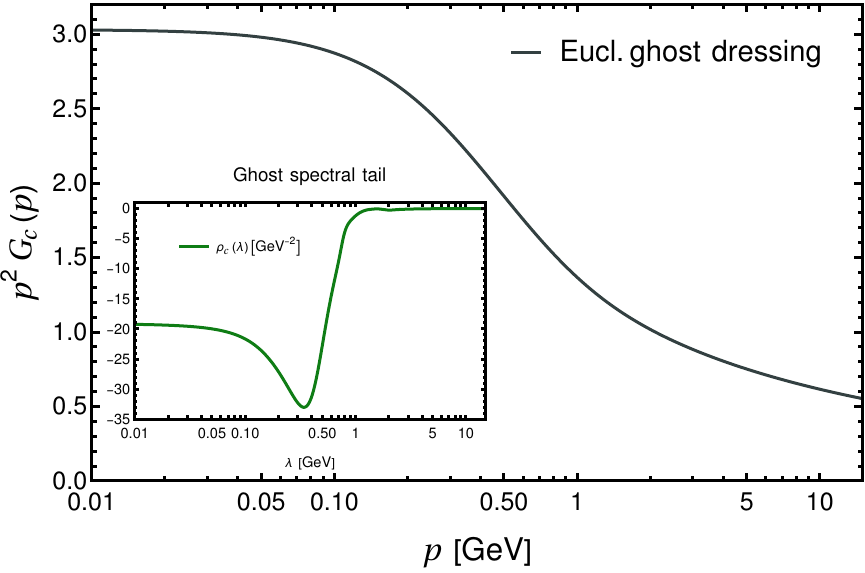}
        \caption{ Euclidean ghost dressing. The corresponding spectral tail is shown in the inset. \hspace*{\fill} \newline }
        \label{fig:ghostdressing}
    \end{subfigure}
    \caption{Input for the spectral gap equation from the reconstructed glue sector. }
    \label{fig:gluesector}
\end{figure*}

For the gluon propagator in (2+1)-flavor QCD, we use the spectral one obtained with a spectral reconstruction in \cite{Horak:2021syv, Horak:2023xfb}. In these works, a Källén-Lehmann representation for the gluon propagator in the Landau gauge was assumed,
\begin{align}
    G_A^{\mu \nu}(q)  = \Pi^{\mu \nu}(q) G_A(q)  \quad \Pi^{\mu \nu}(q) = \delta^{\mu \nu} -\frac{ q^\mu q^\nu}{q^2}\,,
    \label{eq:gluonprop}
\end{align}
with
\begin{align}
    G_A(q) = \int_\lambda \frac{\lambda \rho_A(\lambda)}{q^2 + \lambda^2}\,, \qquad
    \int_\lambda \lambda \rho_A(\lambda) = 0\,.
    \label{eq:gluonspec}
\end{align}
The sum rule in \labelcref{eq:gluonspec} is the Oehme-Zimmermann super-convergence relation \cite{Oehme:1979ai,Oehme:1990kd,Oehme:1994pv} in the Landau gauge. 

For the generality of the analysis of the causal structure of the quark propagator done below, it is important to also discuss the impact of potential cc poles in the gluon propagator, as they have been observed in multiple reconstructions, see eg., \cite{Falcao:2020vyr,Binosi:2019ecz,Siringo:2016jrc}. In the present truncation for the quark-gluon vertex,  they lead to complex conjugate cuts in the quark self energy. However, as the real part of potential complex poles in the gluon propagator would take a value related to the gluon mass scale, their inclusion would not alter the behaviour of the quark spectral function near the threshold. In conclusion, 
the structural results of this work do not depend on the existence or absence of cc poles in the spectral representation of the gluon.

\subsection{A causal quark-gluon vertex}
\label{sec:CausalQuark}

The quark-gluon vertex is the pivotal ingredient in the gap equation and hence for the dynamics of chiral symmetry breaking. It has been thoroughly studied in the literature, see e.g.~\cite{Mitter:2014wpa, Williams:2014iea, Williams:2015cvx, Cyrol:2017ewj, Gao:2021wun, Aguilar:2024ciu}. It can be decomposed into twelve Dirac structures,
\begin{align}
    \Gamma_{\mu}(p,q) = \sum_{i=1}^{12} \lambda^{(i)}(p,q)\,\mathcal{T}^{(i)}_{\mu}(p,q)\,,
\label{eq:quarkvertex}
\end{align}
of which four can be chosen purely longitudinal and hence do not enter the gap equation in the Landau gauge, for basis choices and more details we refer to \cite{Mitter:2014wpa, Cyrol:2017ewj, Gao:2021wun, Aguilar:2024ciu, Ihssen:2024miv}. While we only use the classical tensor structure in the explicit computation, our analysis uses the basis in \cite{Ihssen:2024miv}. We count all momenta as incoming, and  $p$ and $q$ are the incoming gluon and antiquark momenta, respectively. The dominant transverse structures are given by
\begin{align}\nonumber
    \mathcal{T}^{(1)}_{\mu}(p,q) & =\textrm{i} \,\gamma_\mu\,, \qquad
    \mathcal{T}^{(4)}_{\mu}(p,q)  =\textrm{i}\, \sigma_{\mu\nu}p_\nu\,,  \\[2ex] 
    \mathcal{T}^{(7)}_{\mu}(p,q) & =\frac{1}{3}\Bigl[ \sigma_{\alpha \beta}\gamma_\mu +\sigma_{\beta \mu}\gamma_\alpha +\sigma_{\mu \alpha }\gamma_\beta \Bigr]\, k^+_{\alpha} k^-_{\beta} \,,
    \label{eq:T147}
\end{align}
where $\sigma^{\mu\nu}=\textrm{i} /2 [\gamma^\mu,\gamma^\nu]$ and $k^{\pm}=(p\pm q)$.

While the classical tensor structure is the dominant one, ${\cal T}^{(4,7)}$ also have a major impact on the size of chiral symmetry breaking. Specifically, the chiral symmetry breaking tensor structure ${\cal T}^{(4)}$ plays a sizeable rôle in the generation of the constituent quark mass in the chirally broken regime. Its impact on the causal structure will be studied in a forthcoming work, but here we focus on the analytic structure induced by the overall strength of the classical Dirac structure. Apart from being the dominant effect and hence deserves to be studied first, respective vertices have been and are used in manifold applications in low energy QCD, and are common place in bound state studies: roughly speaking, in all these applications one drops ${\cal T}^{(4,7)}$ and emulates the missing fluctuations by increasing the infrared strength of the dressing $\lambda^{(1)}$ of the classical tensor structure.

Hence, we proceed likewise here by only using the first tensor structure, where we identify the corresponding form factor with the ghost dressing to capture the UV running and peak position of the quark-gluon vertex. This dressing can be inferred by the Slavnov-Taylor identity and resembles a simplified variant of the Ball-Chiu vertex \cite{Ball:1980ay}. The STIs can also be used to construct a fully gauge-consistent causal vertex, see \Cref{sec:Slavnov-Tailor}. For a comparison of STI vertices with the full ones, see e.g.~\cite{Gao:2021wun}. The vertex is then given by
\begin{subequations}
  \label{eq:vertexdressing1}
  \begin{align}
    \Gamma_{\mu}(p,q) = \textrm{i}\, \gamma_\mu  \lambda_1(p^2)\,,
      \label{eq:vertexdressing1A}
  \end{align}
  with
  \begin{align}
      \lambda_1(p^2)=\eta\, g_s(\mu_r) \frac{Z_q(\mu_r^2)}{Z_c(p^2)}\,.
      \label{eq:vertexdressing1B}
  \end{align}
\end{subequations}
The form factor $\lambda_1(p^2)$ contains an additional factor $Z_q(\mu_r)$ with the renormalisation group scale $\mu_r$ for RG-consistency and a global strength factor $\eta$. Note, that the global strength factor $\eta$ accommodates two qualitatively different contributions: firstly, it compensates for the additional terms in the STI that have been dropped for the sake of convenience and numerical simplicity in the construction \labelcref{eq:vertexdressing1B}. Second, it compensates for dropping the fluctuation effects from the other vertex structures, and in particular  ${\cal T}^{(4,7)}$. For other constructions of the quark-gluon vertex in the context of bound state studies and the Dyson-Schwinger equation, see \cite{Bender:2002as,Fischer:2003rp,Bhagwat:2004hn,Fischer:2005en,Fischer:2006ub,Fischer:2007ze,Fischer:2008wy,Chang:2009zb,Chang:2011ei,Fischer:2012vc,Aguilar:2018epe,Gao:2024gdj}.

In the present work, we use the spectral representation of the ghost dressing function in \labelcref{eq:vertexdressing1B}. It has been calculated directly in \cite{Horak:2021pfr} and this computation reveals its spectral structure. This facilitates and constrains its spectral reconstruction from Euclidean precision data, as done in \cite{Horak:2021syv,Horak:2023xfb}. Since the ghost propagator contains a massless pole, the spectral function has to contain a delta function at zero. The spectral representation of the ghost dressing function $1/Z_c(p^2)=p^2G_c(p)$ reads
\begin{align}
    \frac{1}{Z_c(p^2)} = \frac{1}{Z_c(0)}+ p^2\int_{0}^{\infty} \frac{d \lambda}{\pi} \frac{\lambda\,\rho_c(\lambda)}{p^2 + \lambda^2}\,.
    \label{eq:ghostspec}
\end{align}
While the residue of the massless pole is determined by the value of the ghost dressing function at zero, the spectral function $\rho_c(\lambda)$ can be quantitatively described by a rather simple functional form as discussed in \cite{Horak:2021pfr}.
The ghost dressing function and the corresponding spectral function are shown in \Cref{fig:ghostdressing}. It enters the gap equation only as the product of the ghost dressing and the gluon propagator $\bar G_A(p)=G_A(p)/Z_c(p)$, with
\begin{align}
    \bar{G}_A(p) = \int\limits_{0}^{\infty} \frac{d \lambda}{\pi} \frac{\lambda \,\bar \rho_A(\lambda)}{q^2 + \lambda^2}\,.
    \label{eq:barrhoA}
\end{align}
\begin{figure*}[ht]
    \centering
    \begin{subfigure}{.48\linewidth}
        \centering
        \includegraphics[width=.99\textwidth]{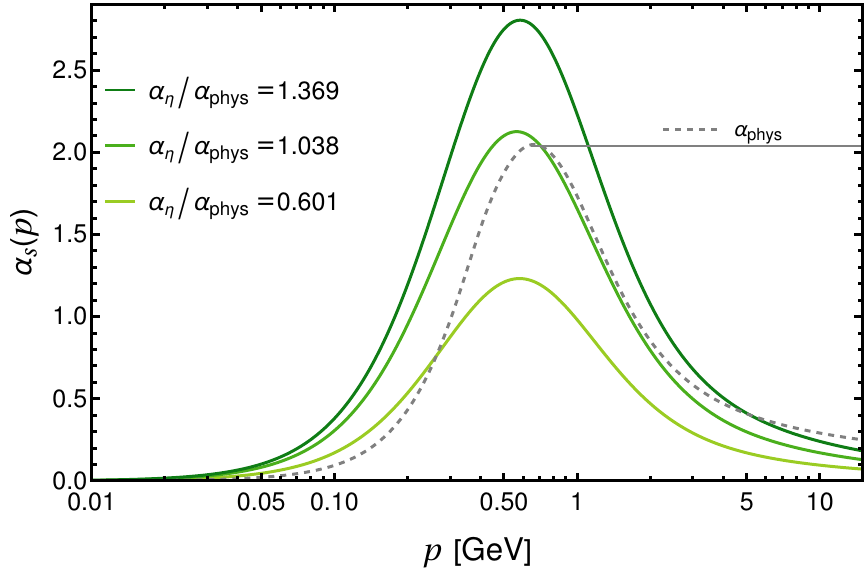}
        \caption{Strong coupling $ \alpha_s$ as a function of momentum and enhancement factor. The grey line indicates the peak value of the full quark-gluon coupling on the symmetric point for $(2+1)$ flavors from \cite{Ihssen:2024miv}. \hspace*{\fill}}
        \label{fig:alpha_s}
    \end{subfigure}
    \hspace*{.51cm}
    \begin{subfigure}{.48\linewidth}
        \includegraphics[width=.99\textwidth]{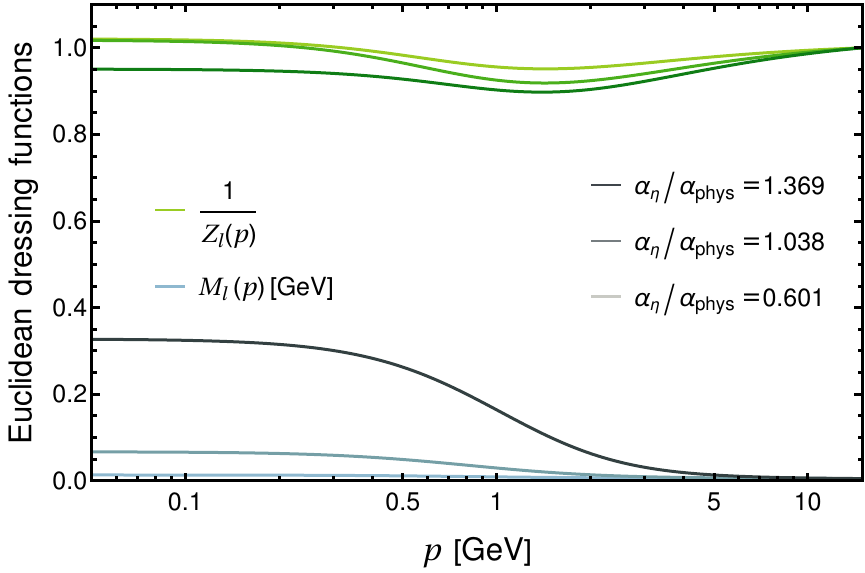}
        \caption{Dressing of the light quark propagator $1/Z_l(p)$ and its mass function $M_l(p)$ for different enhancement factors. The mass function is renormalised to $m_l=2.7$ MeV at the renormalisation scale.\hspace*{\fill}}
        \label{fig:MZplot}
    \end{subfigure}
    \caption{Couplings $\alpha_\eta(p)$ and light quark dressings $M_l(p), 1/Z_l(p)$ for different enhancement factors. }   
    \label{fig:EuclideanPlots}
\end{figure*}
The spectral function $\bar\rho_A(p)$ is shown in \Cref{fig:gluondressing} in comparison to that of the gluon, $\rho_A(p)$. In the present work, we use the spectral reconstruction of the lattice-type decoupling solution from \cite{Horak:2023xfb}.

We also note in passing, that due to its well-understood spectral form one may simply use the fitting form of the ghost spectral function devised in \cite{Horak:2021pfr}. This can be used to model the form factor of the classical tensor structure, although a cautious monitoring of the complex structure of the model is required for minimising the impact of unphysical approximation artefacts.

In summary, the STI construction in \labelcref{eq:vertexdressing1} allows us to reproduce the form of the quark-gluon coupling in the IR. This coupling is defined via the exchange process of a gluon between two quarks projected on the classical tensor structure of the vertex on the symmetric point,
\begin{align}
    \alpha_s(\bar p)= \frac{1}{4 \pi} \frac{\left(\lambda^{(1)}(\bar p)\right)^2}{Z_q(\bar p)^2 Z_A(\bar p)}\,, \qquad \bar p^2 = \frac13 \sum_{i=1}^3 p_i^2\,.
    \label{eq:alphaqugl}
\end{align}
We will use results for \labelcref{eq:alphaqugl} from quantitative functional computations in \cite{Cyrol:2017ewj, Gao:2021wun, Ihssen:2024miv} as a point of reference, as they have been computed in the renormalisation scheme also employed here: In our computation, both the gluon and ghost dressings are fixed by the inputs; see \Cref{fig:gluondressing} and \Cref{fig:ghostdressing}. Accordingly, \labelcref{eq:alphaqugl} only depends on the global strength factor $\eta$ introduced in \labelcref{eq:vertexdressing1B} and the $\eta$-dependent quark dressings. To compare against the physical coupling, we define the respective peak values as
\begin{align}
    \alpha_{\textrm{\tiny{phys}}} = \max_{\bar p} \alpha_s(\bar p)\,, \qquad
    \alpha_\eta = \max_{\bar p} \alpha^{\eta}_s(\bar p)\,.
\label{eq:alphapeaks}
\end{align}
For the physical coupling, we use the result of the quantitative computation in \cite{Ihssen:2024miv}. The superscript $\eta$ indicates the dependence of the coupling used in the present work on the global enhancement factor $\eta$. This dependence comes explicitly from the vertex dressing \labelcref{eq:vertexdressing1B} and implicitly from the quark dressing function, which we determine dynamically from the gap equation for a given value of $\eta$. We will use the ratio $\alpha_\eta/\alpha_{\textrm{\tiny{phys}}}$ to quantify the enhancement of the coupling due to the global strength factor $\eta$. The couplings and the Euclidean quark dressing functions are given in \Cref{fig:alpha_s} and \Cref{fig:MZplot} respectively.
In \Cref{fig:alpha_s} we show the exchange coupling for various global strength factors $\eta \in \lbrace 0.553, 0.742,0.883\rbrace$ in comparison to the physical value. The corresponding rations of the peak are $\alpha_\eta/\alpha_{\textrm{\tiny{phys}}} \in \lbrace 0.601,1.038,1.369\rbrace$. We note that the approximate vertex \labelcref{eq:vertexdressing1} reproduces the qualitative features of the exchange couplings well, in particular the peak position: it is located at about $p_{\textrm{\tiny{glue}}}^{\textrm{\tiny{peak}}}\approx 0.6$\,GeV, close to that of the full solution in \cite{Cyrol:2017ewj, Gao:2021wun, Ihssen:2024miv}. The physical amount of chiral symmetry breaking requires a strong enhancement of the coupling with $\alpha_\eta/\alpha_{\textrm{\tiny{phys}}} \approx1.4$. However, part of this enhancement is also related to the details of the coupling. For the total amount of chiral symmetry breaking, it is rather the integrated coupling that is relevant. This includes in particular also the steepness of the coupling on both sides of the peak, which is not captured by our vertex construction.

\begin{figure*}[ht]
    \centering
    \begin{subfigure}{.48\linewidth}
        \centering
        \includegraphics[width=.99\textwidth]{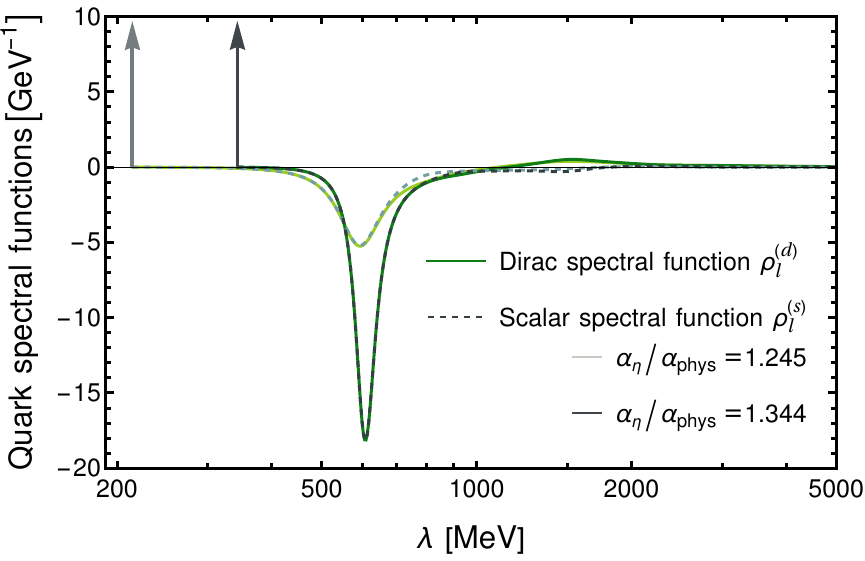}
        \caption{Quark spectral functions for different enhancement factors, which correspond to qualitatively different complex structures. The arrows indicate real poles at the onset of the scattering continuum with $m_\textrm{\tiny{pole}}\in \lbrace 214, 342 \rbrace $ MeV. \hspace*{\fill}}
        \label{fig:rhoplot}
    \end{subfigure}
    \hspace*{.51cm}
    \begin{subfigure}{.48\linewidth}
        \vspace{0.1cm }
        \includegraphics[width=.99\textwidth]{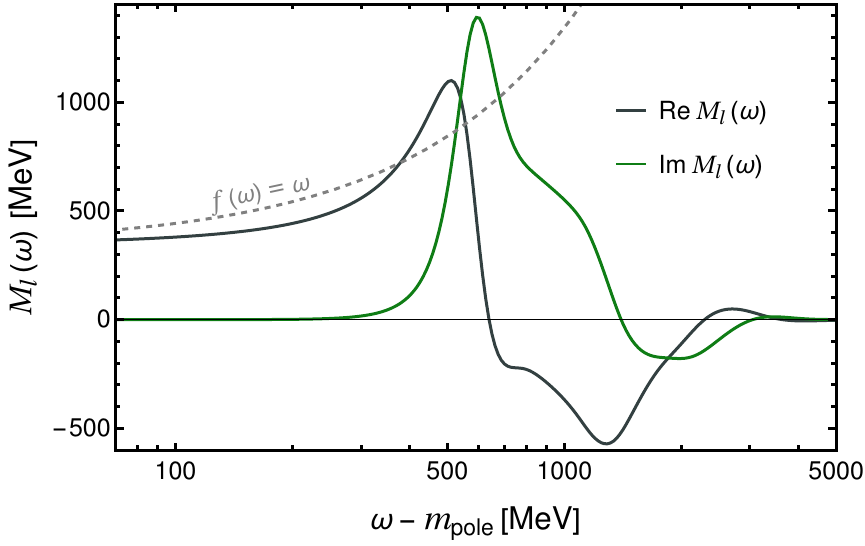}
        \caption{Real and imaginary parts of the mass function for \mbox{$\alpha^\eta/\alpha_{\textrm{\tiny{phys}}}=1.344$}, see \labelcref{eq:vertexdressing1B,eq:alphapeaks}. The x-axis shows $\omega - m_\text{pole}$. The real part crosses the line $f(w)=w$, indicated in grey, what signals cc poles. \hspace*{\fill}}
        \label{fig:ReImMplot}
    \end{subfigure}
    \caption{Spectral functions $\rho_{l}^{(d,s)}$ of the quark propagator (left), and its mass function $M_l$ for real frequencies (right). }
    \label{fig:ComplexStructure}
\end{figure*}

\section{The causal structure of the quark propagator}
\label{sec:Results}

The spectral gap equation is solved numerically with the input discussed in the last section. Inserting the quark-gluon vertex \labelcref{eq:vertexdressing1} in the gap equation \labelcref{eq:selfenergy} and using the spectral representations of the quark propagator  \labelcref{eq:quarkspecvac} and that of the scattering kernel $G_A(p)/Z_c(p)$ in \labelcref{eq:barrhoA}, we arrive at
\begin{align}
    \Sigma_q(p) =    g_s^2\, C_f Z_1  \!\!\int\limits_{\lambda_q,\lambda_A}\!\!\!\!  \lambda_A\,\bar{\rho}_A(\lambda_A) \rho_q(\lambda_q) I(\lambda_q,\lambda_A,p)\,,
    \label{eq:spectralselfenergy}
\end{align}
with
\begin{align}
    \ I(\lambda_q,\lambda_A,p) = \int_q \frac{\gamma_\mu \Pi^{\mu \nu}(q)\left( -\textrm{i} (\slashed p + \slashed q) + \lambda_q\right)\gamma_\nu  }{\left(q^2 + \lambda_A^2\right)\left((p+q)^2 + \lambda_q^2\right)}\,.
    \label{eq:DefofI}
\end{align}
The loop integral \cref{eq:DefofI} is computed in dimensional regularisation, and the whole self energy is renormalised within a BPHZ subtraction scheme,
\begin{align}\label{eq:sigmaren}
    \Sigma^\textrm{\tiny{ren}}_q(p) = \Sigma_q(p) - \Sigma_q(\mu_r) \,,
\end{align}
where the subtraction implicitly determines the renormalisation constants $Z_2$ and $Z_m$ in \cref{eq:quarkDSE}. This procedure not only removes the explicitly divergent terms of the loop integral \cref{eq:DefofI}, but also the remaining divergences from the spectral integral. For details on the spectral renormalisation, see \cite{Horak:2020eng} and for the explicit implementation in the present work, see \Cref{sec:selfenergy-calculations}.

The spectral function of the scattering kernel $\bar\rho_A(\lambda)$ is shown in \Cref{fig:gluondressing}. Since the vertex dressing does not diverge in the infrared, the combined spectral function still approaches zero at the onset of the scattering continuum fast enough and $\bar{G}_A$ remains finite at the origin. 
Moreover, the imaginary part of both, the Dirac and scalar part of $I(\lambda_q,\lambda_A,p)$ in \labelcref{eq:DefofI}, contains only terms proportional to either $\sqrt{w-\lambda_q -\lambda_A}\,\Theta(w-\lambda_q -\lambda_A)$ or  $(\omega - \lambda_q)^3\,\Theta(w-\lambda_q)$, see \Cref{sec:selfenergy-calculations} for details. Consequently, the imaginary part of the self energy vanishes at its onset with a power law. Hence, the self-consistent iteration converges towards a solution with a first zero crossing in $ w- \text{Re}\,M_l(w)$ that coincides with a vanishing imaginary part of the mass function. This leads to a real pole as the first singularity of the quark propagator in the complex plane, which is discussed in detail in the following \Cref{sec:analyticStructure}. In \Cref{sec:CCPoles} we complete the discussion of the causal structure of the quark propagator with that on the emergence of cc poles, in addition to the real one, for large coupling strength.

\subsection{The real pole of the quark propagator}
\label{sec:analyticStructure}

For the evaluation of the qualitative change of the causal structure, we scan the strength parameter $\eta$ in the vertex  \labelcref{eq:vertexdressing1} in the interval $[0.553,0.883]$. This corresponds to a relative coupling strength $\alpha_\eta/\alpha_{\textrm{\tiny{phys}}}$ in the interval $[0.601,1.36941] $. Technically, it is convenient to solve the spectral DSE first for the smallest value of $\eta$ considered here and use the result iteratively as the input for the first iteration of the gap equation for the next bigger $\eta$-value.

We exemplify our results for the spectral functions in \Cref{fig:rhoplot}, where $\rho_{l}^{(s,d)}$ of the light quarks are shown (left panel) for two different values of the vertex strength $\eta= 0.829, 0.872$ with different causal structures. These $\eta$'s correspond to relative coupling strengths $\alpha_\eta/\alpha_{\textrm{\tiny{phys}}} = 1.245, 1.344$. The real pole at $w=m_\textrm{\tiny{pole}}$ is indicated by a delta peak in the spectral functions in \Cref{fig:rhoplot}. The cut, induced by the gluon exchange, starts at the pole, but is vanishing there. The imaginary parts of mass function and wave function both approach zero at the onset, see \Cref{fig:ReImMplot} for the mass function. We would like to emphasise that a real pole in the quark propagator does not correspond to a physical, asymptotic particle state that could be observed in the experiment. For larger values of the spectral parameter, the spectral functions are predominantly negative. This positivity violation is necessary for satisfying the spectral sum rules in \labelcref{eq:sum rules}. In the right panel, \Cref{fig:MZplot}, we show the mass function $M_l(\omega)$ for the larger value ${\alpha_\eta/\alpha_{\textrm{\tiny{phys}}} = 1.344}$: the real part of the mass function crosses the line $f(\omega)=\omega$, indicating the emergence of additional pairs of cc poles, which will be discussed in \Cref{sec:CCPoles}. 

The most prominent common feature of the spectral functions for all tested coupling strengths is the emergence of a real pole as the first singularity in the complex plane at the onset of the scattering continuum. Evidently, we cannot draw fully conclusive statements about the nature of the first singularity in the complex plane, as the vertex is not computed self-consistently. Instead, we use an STI-compatible vertex model. However, the mechanism that underlies the occurrence of the real pole is rather generic. It is simply based on two properties of the product of the gluon propagator and the vertex, which we expect to hold in full QCD in the Landau gauge. We shall discuss these properties by comparing our result with the results of direct real time computations in the literature: in most cases, including the present work, the effective quark-gluon interaction is modelled by an effective gluon exchange that combines the gluon propagator and the vertex,
\begin{align}\label{eq:effectiveinteraction}
    G_A(q)\Gamma_{\mu}(p,q) \propto \gamma_\mu \frac{\alpha_s^{\textrm{\tiny{eff}}}(q)}{q^2}\,.
\end{align}
The first property is the finiteness of \labelcref{eq:effectiveinteraction} in the soft gluon limit. A singularity such as a pole, a branch-cut singularity, or a logarithmic divergence, signalled by a finite value of a cut at vanishing gluon momentum, possibly shifts the first singularity into the complex plane. Previous results that show a complex first singularity can be found in \cite{Stainsby:1990fh,Maris:1991cb,Fischer:2008sp} for QCD-like models in four dimensions and \cite{Maris:1995ns} in QED$_3$. While the details of these vertex models differ, the combination of the gluon propagator and vertex diverges in the soft gluon limit. 
In \cite{Stainsby:1990fh,Maris:1995ns}, the effective coupling $\alpha^s_\textrm{eff}$ stays finite in the soft gluon limit, effectively leading to the implementation of a massless gluon (or photon) propagator. The same holds true for \cite{Fischer:2008sp}, where the authors use a similar construction to that of the present work, only that the dressing function $\lambda_1$ is proportional to $1/Z_c(p)^2$. This leads to an $\alpha_s^{\textrm{\tiny{eff}}}$ that is proportional to the Taylor coupling, which acquires a finite value on the scaling solution. The confinement term in \cite{Maris:1991cb} even contains a delta function at the origin. In full computations, the product of the gluon propagator and the vertex stays finite in the soft gluon limit, see \cite{Mitter:2014wpa, Williams:2014iea, Williams:2015cvx, Cyrol:2017ewj, Gao:2021wun}. Note in this context, that this property is fulfilled in full calculations even if considering explicitly the dynamical generation of confinement by means of the Schwinger mechanism, see e.g.~\cite{Aguilar:2011xe, Aguilar:2021uwa, Ferreira:2023fva, Aguilar:2022thg}, or the quartet mechanism, see e.g.~\cite{Alkofer:2011pe}. We discuss this at the example of the Schwinger mechanism. There, the confining mass gap of the gluon in the Landau gauge is related to the occurrence of a massless bound state in the longitudinal glue sector of the theory. While this massless bound state is mirrored in the longitudinal part of the quark-gluon vertex, the transverse quark-gluon vertex does not contain a pole at vanishing gluon momentum.

The second assumption that appears to be crucial for our analysis to hold, is the existence of a Källén-Lehmann representation for the product of the gluon propagator and the vertex. While the standard Rainbow-Ladder kernel \cite{Maris:1999nt} does not have a singularity in the origin, it also does not admit a spectral representation due to the finite width representation of $\delta(q)$. The latter implements the necessary infrared enhancement of the vertex but also carries an essential singularity at complex infinity. The nature of the first singularity in the complex plane for such models depends on the explicit parameters and has been studied in, e.g., \cite{Windisch:2016iud} and \cite{Eichmann:2009zx}, where the dependence of hadron masses on the position of the first pole is found to be mild in the latter.

Further studies that link the absence of a real pole to confinement, such as \cite{Bhagwat:2003vw,Alkofer:2008tt}, only deduce the possibility of cc poles from the positivity violation of the Schwinger function, which does not rule out a real pole, whereas the reconstructions in \cite{Alkofer:2003jj,Falcao:2022gxt} indicate the existence of a dominant singularity on the real axis. The scenario of a real pole was also found in \cite{Solis:2019fzm} for a vertex model that resembles a massive gluon with the mass $m_\textrm{\tiny{glue}}$, i.e., where the cut of the quark self energy starts only at $m_\textrm{\tiny{pole}} + m_\textrm{\tiny{glue}}$, different to our construction. Moreover, the authors also report the emergence of additional cc poles for certain parameter choices, without giving a detailed analysis of the underlying mechanism.

\begin{figure*}[ht]
    \centering
    \begin{subfigure}{.48\linewidth}
        \centering
        \includegraphics[width=.94\textwidth]{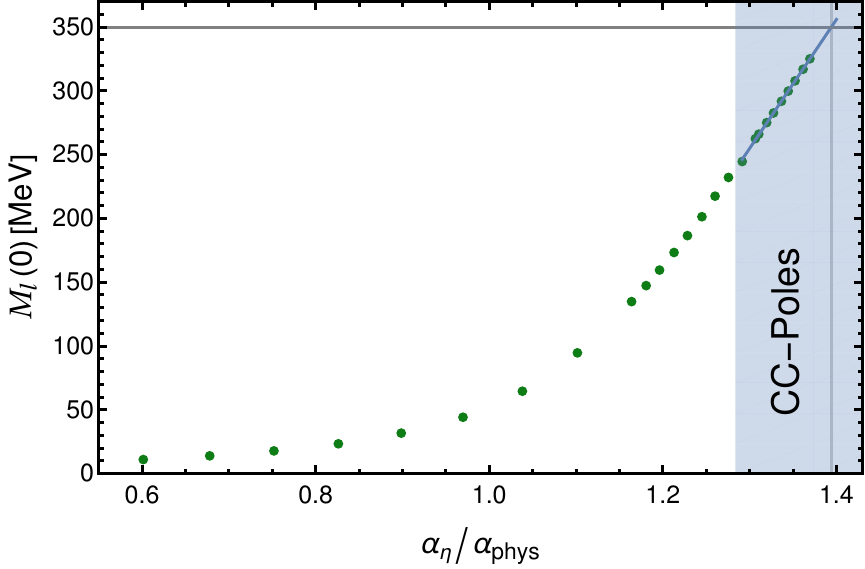}
        \caption{Constituent quark mass over the enhancement factor. We find the emergence of cc poles for larger enhancement factors, that are indicated by the blue region. The grey lines indicate an estimate of the necessary enhancement to achieve a realistic amount of D$\chi$SB with $M(0)\approx 350$ MeV  \hspace*{\fill}}
        \label{fig:M0overenhancement}
    \end{subfigure}
    \hspace*{.31cm}
    \begin{subfigure}{.48\linewidth}
        \includegraphics[width=.99\textwidth]{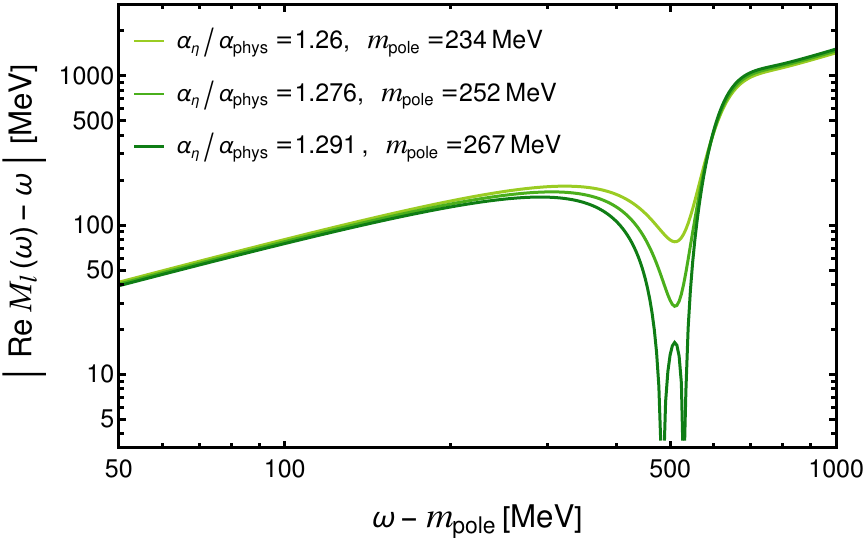}
        \caption{The absolute value of the $\text{Re }M(\omega)-w$ for different values of the coupling. For the largest value, the zero crossings show the emergence of cc poles. The critical value of the coupling is approximately $\alpha_{\eta^*}/\alpha_{\textrm{\tiny{phys}}} \approx 1.283$ \hspace*{\fill}\newline}
        \label{fig:AbsMminwplot}
    \end{subfigure}
    \caption{Emergence of the complex conjugate poles for larger vertex strengths.}
    \label{fig:EuclideanfunctionCCpoles}
\end{figure*}
%

\subsection{Additional cc poles for large vertex strength}
\label{sec:CCPoles}

Having a real pole established as the first singularity of the quark propagator in our approximation, we proceed with the analysis of the remaining causal structure. The current approximation reproduces D$\chi$SB quantitatively, subject to an appropriate choice of the strength parameter $\eta$. The physical constituent quark mass is obtained for 
\begin{align}
M_l(0) \approx 350\,\textrm{MeV},\quad \eta_{\textrm{\tiny{const}}}=0.882: \ \frac{\alpha_\eta}{\alpha_{\textrm{\tiny{phys}}}}\approx  1.39, 
\label{eq:MLphys} 
\end{align}
see \Cref{fig:M0overenhancement}. The ratio $\alpha_\eta/\alpha_{\textrm{\tiny{phys}}}$ in \labelcref{eq:MLphys} is the enhancement of the peak coupling for a given $ \eta_{\textrm{\tiny{const}}}$. This is a good measure for the overall enhancement of the vertex strength. We shall see that in this case the quark propagator has no Källén-Lehmann representation but contains at least two cc poles in addition to the real one.

In turn, using the vertex strength with the physical peak height of the coupling, the respective constituent quark mass is given by 
\begin{align}
    M_l(0) \approx 60\,\textrm{MeV},  \quad \eta_{\textrm{\tiny{phys}}}=0.741: \ \frac{\alpha_\eta}{\alpha_{\textrm{\tiny{phys}}}}\approx  1, 
    \label{eq:ML-alphaphys} 
\end{align}
see \Cref{fig:EuclideanPlots}. This compares well with the solution of the gap equation by using the full dressing $\lambda^{(1)}$ in \cite{Gao:2021wun} with $M_l(0)\approx 80$\,MeV, the difference can be attributed to the slightly different shape of the vertex functions, as the width and ultraviolet decay of $\lambda^{(1)}$ have a sizeable impact on the strength of chiral symmetry breaking. In any case, this illustrates impressively the importance of the other tensor structures, see e.g.~\cite{Mitter:2014wpa, Williams:2014iea, Williams:2015cvx, Cyrol:2017ewj, Gao:2021wun}, see also the review \cite{Eichmann:2016yit}.  
The full constituent quark mass is obtained with the physical vertex strength if also ${\cal T}^{(4,7)}$ are taken into account, a rather comprehensive analysis is given in \cite{Gao:2021wun}.

In short, while the constituent quark mass is significantly smaller than the physical one for the physical vertex strength with $\eta_{\textrm{\tiny{phys}}}$, the corresponding quark propagator has a Källén-Lehmann representation. This has to be contrasted with the results for $ \eta_{\textrm{\tiny{const}}}$ with the physical constituent quark mass, where we end up with cc poles.

We proceed by dissecting the emergence of these cc poles, including a preliminary assessment of the situation and possible causal structure of the quark propagator in full QCD, where we take into account the additional tensor structures of the quark-gluon vertex:

In \Cref{fig:M0overenhancement}, we show the constituent quark mass as a function of the strength factor $\eta$, or rather the respective ratio $\alpha_\eta/\alpha_{\textrm{\tiny{phys}}}$. The respective (2+1)-flavor quark-gluon coupling is taken from \cite{Ihssen:2024miv}. Our computation shows the existence of a critical value of the vertex strength,
\begin{align}
    \eta^*\approx 0.846\,,\qquad \longleftrightarrow \qquad  \frac{\alpha_{\eta^*}}{\alpha_{\textrm{\tiny{phys}}}}\approx 1.283\,.
    \label{eq:eta*}
\end{align}
For $\eta<\eta^*$, the quark propagator has a Källén-Lehmann representation, and the physical vertex strength is located in this regime.
For $\eta>\eta^*$, we keep the real pole as the first singularity in the complex plane, but additional complex conjugated poles emerge. The enhancement of the vertex strength required to obtain the physical constituent quark mass $M_l(0) \approx 350$\,MeV in the current approximation is given by $\alpha_\eta/\alpha_{\textrm{\tiny{phys}}} \approx 1.39$ and is deep in the regime with cc poles. It is indicated by the vertical straight grey line in \Cref{fig:M0overenhancement}.

It is left to assess the origin of the cc poles, which allows us to discuss its presence or absence in full QCD. The occurrence of the cc poles originates in the analytic structure of the mass function and is connected to the height and width of the effective coupling. The peak structure imprints itself on the imaginary parts of the mass function, see \Cref{fig:ReImMplot}, and leads to a peaked structure with zero crossing in the real part. The cc poles emerge, when the peak of the real part crosses the line $f(\omega)=\omega$, i.e., when $\text{Re}\,M(\omega) - \omega = 0$, see \Cref{fig:AbsMminwplot}. In comparison to the computation with a constant vertex dressing in \cite{Horak:2022aza}, this novel feature is connected to the non-trivial momentum dependence of the quark-gluon vertex. This dynamical property shifts the peak of the effective coupling relevant for the solution of the gap equation to a more realistic value. The peak position in the full calculations are approximately $590$\,MeV for two flavors, see \cite{Cyrol:2017ewj} and $670$\,MeV, \cite{Ihssen:2024miv} for (2+1) flavors. The boundary of the regime with  cc poles certainly depends on the details of the vertex. However, it is dominantly sensitive to the position, width, and height of the quark-gluon coupling, as these properties influence the height and position of the peaked structure in the real part of the mass function significantly. Our results suggest, that for physical peak position of the coupling in two- or $(2+1)$-flavor QCD, cc poles are easily introduced by an enhancement of the vertex. 

In summary, it is suggestive that the absence of cc poles seen for a physical vertex strength in the present approximation persists in full QCD: the inclusion of the other tensor structures may change the peak structure of the mass function, leading to the absence of cc poles. Indications for this scenario are present in the full vertex solutions, e.g., \cite{Cyrol:2017ewj, Gao:2021wun}. There, the different momentum running of the dominant couplings $\lambda^{(4)}$ and $\lambda^{(7)}$ lead to different peak positions in the associated effective couplings. Hence, the inclusion of these contributions might simply lead to a broadened structure of the imaginary part of the mass function and allows for a larger constituent quark mass without the emergence of cc poles. Note, that while we consider the above scenario as the likely one, we can not make a definite statement about the causal structure with a full vertex prior to a direct computation. This matter will be resolved in a forthcoming work, where a computation with the physical quark-gluon vertex with the tensor structures ${\cal T}^{(1,4,7)}$ is performed. 

\begin{figure}[t]
    \centering
    \includegraphics[width=.45\textwidth]{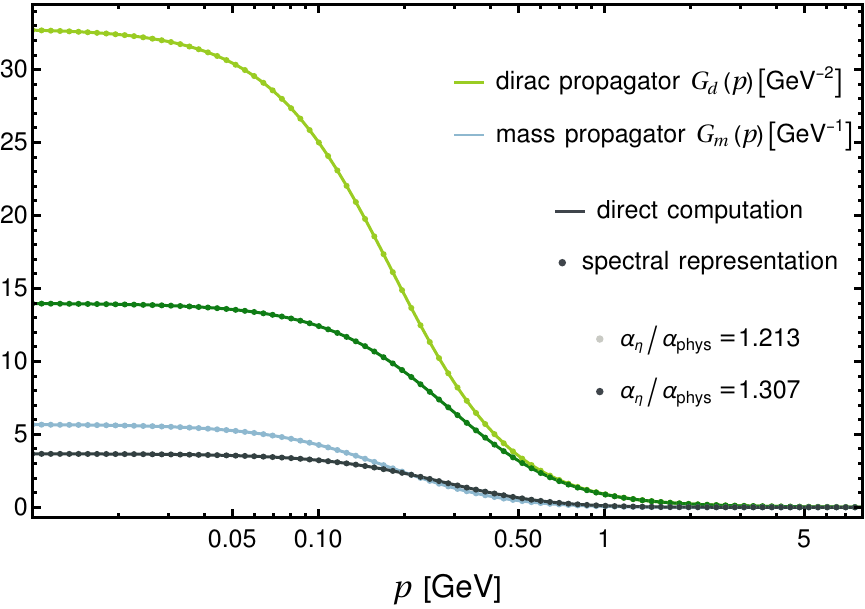}
    \caption{Propagators in the Euclidean domain for different enhancement factors. The spectral representation is fulfilled up to the permille level for all coupling strengths, where the higher enhancement factor already has cc poles.\hspace*{\fill}}
    \label{fig:euclcrosscheck}
\end{figure}
We close this section with a discussion of the relevance of the findings above for existing bound state studies. To that end, we remark, that the additional cc poles seem to only have a mild influence on the Euclidean correlation functions. As can be seen in \Cref{fig:euclcrosscheck}, the Källén-Lehmann spectral representation holds true up to the permille level also for coupling strengths where the cc poles are present in the system. The two pairs of complex poles have residues of opposite signs, see \Cref{fig:AbsMminwplot}, what leads to cancellations. This is particularly interesting for the applications to hadron physics. There, most numerical and direct analytical continuation methods rely on simple causal structures, and the presence of cc poles could have serious consequences for the validity of expansion schemes or spectral representations. With such applications in mind, we would like to emphasise, that the cc poles are, if present, rather far away from the real frequency axis. Their location is tied to the peak-position of the imaginary part of the mass function. Moreover, the absolute distance of the cc poles to the origin is between $0.6-1$\,GeV or even larger. This originates in the fact that their location in the quark mass function is approximately given by the sum of the quark pole mass and the peak position of the gluon dressing,  $m_\textrm{\tiny{pole}} + p_{\textrm{\tiny{glue}}}^{\textrm{\tiny{peak}}}$. For applications that do not probe this region in the complex frequency plane, as for example the computation of light hadron properties, the effects of these cc poles may even be negligible.

\section{conclusion}
\label{sec:conclusion}

In the present work, we have studied the causal structure of the quark propagator within the spectral DSE. In contrast to the previous work \cite{Horak:2022aza}, we made a significant step towards its solution in full QCD: instead of a constant quark-gluon vertex we have used an STI-construction for the vertex, that includes a fully momentum-dependent dressing for the classical tensor structure, see \Cref{sec:CausalQuark}. This allowed us to study the complex structure of the quark propagator by solving the spectral gap equation.

Specifically, we have shown in \Cref{sec:Results} that the quark propagator has a Källén-Lehmann representation for the physical vertex strength of the classical tensor structure, but fails to generate the correct strength of spontaneous chiral symmetry breaking. In turn, enhancing the quark-gluon coupling above the physical value with a factor $1.39$ leads to the correct amount of chiral symmetry breaking. This procedure is the commonly used one for DSE and bound state studies, but introduces cc poles to the quark propagator in addition to the mass like poles that are associated with the first singularity in the complex plane. The nature of the latter cannot be answered conclusively without a direct computation with a fully frequency dependent vertex. However, we have discussed the relevant properties of common vertex approximations that decide about a real or complex pole scenario in \Cref{sec:analyticStructure}. Independent of the nature of the first pole, our discussion in \Cref{sec:CCPoles} reveals that the use of vertex models with an over-enhancement of the vertex generically lead to additional cc poles. Importantly, our results also entail that, although current vertex models used in DSE and bound state applications lead to cc poles in the quark propagator, the contribution of the latter to the propagator is negligible. Hence, effectively the quark still has a Källén-Lehmann representation. 

A dissection of the mechanisms suggests that in full QCD, the additional relevant tensor structures in the quark-gluon vertex may lead to the full strength of chiral symmetry breaking without the additional cc poles. This assessment is based on the absence of these cc poles for the physical vertex strength of the classical tensor structure and the momentum dependence of these dressings, see e.g.,~\cite{Cyrol:2017ewj}. The full computation including the additional tensor structures is the subject of ongoing work, and we hope to report on the respective results in the near future.

\begin{acknowledgments}

We thank Gernot Eichmann, Christian Fischer, Joannis Papavassiliou and Franz Sattler for discussions and collaborations on related projects. This work is done within the fQCD-collaboration~\cite{fQCD}, and we thank the members for discussion and collaborations on related projects. This work is funded by the Deutsche Forschungsgemeinschaft (DFG, German Research Foundation) under Germany's Excellence Strategy EXC 2181/1 - 390900948 (the Heidelberg STRUCTURES Excellence Cluster) and under the Collaborative Research Centre SFB 1225 (ISOQUANT).

\end{acknowledgments}

\appendix

\section{The spectral STI-vertex}
\label{sec:Slavnov-Tailor}

The Slavnov-Taylor identity (STI) of the Quark-Gluon vertex reads
\begin{align}\nonumber
     k_\mu \Gamma^\mu_{\bar{q}qA}(k,p)  =  &\,  g  \frac{1 }{ Z_c(k)}t^a\times  \\[2ex]
      \Bigl[ G_q^{-1}(p+k)& {\cal H}(k,q) -G_q^{-1}(p)\bar{\cal H}(-k,p+q)\Bigr]\,,
\label{eq:QGV-STI}
\end{align}
for a derivation, see, e.g., \cite{Cyrol:2017ewj}. The inclusion of non-trivial scattering kernels goes beyond the scope of the present work. We drop it in an Abelian approximation of the STI, ${ \cal H}(k,q)=1$,  and only keep the prefactor $g t^a$ on the right-hand side together with the overall ghost dressing function $1/Z_c(k)$. The latter leads to an enhancement of the vertex in the IR. For a non-abelian computation see \cite{Aguilar:2018epe}. The inverse quark propagators are cancelled by multiplying \labelcref{eq:QGV-STI} from both sides with the respective quark propagator, also using their spectral representation,   
\begin{align}\nonumber
     & k_\mu G_q(p+k)\, \Gamma^\mu_{\bar{q}qA}(k,p)\,G_q(p) \\[2ex]\nonumber
     & = g_s  \frac{1}{Z_c(k)} t^a \Bigl[ G_q(p+k)-G_q(p)\Bigr] \\[2ex]
     & =-\textrm{i}  g_s  \frac{1}{Z_c(k)} t^a\int_\lambda \rho_q(\lambda) \,\frac{1}{\textrm{i}(\slashed p+\slashed k) +\lambda} \,\slashed k\, \frac{1}{\textrm{i}\slashed p + \lambda}\,. 
    \label{eq:SimplifiedSTI}
\end{align}
\Cref{eq:SimplifiedSTI} admits the simple solution, 
\begin{align}\nonumber
      G_q(p+k) \Gamma^\mu_{\bar{q}qA}(k,p)G_q(p) \approx -\textrm{i}  g_s  \frac{1}{Z_c(k)} t^a {\cal K^{\mu}}(k,p)(p)\,,
\label{eq:SimplifiedSTI-Vertex}    
\end{align}
with
\begin{align}
    {\cal K^{\mu}}(k,p)  = \int_\lambda \rho_q(\lambda) \frac{1}{\textrm{i}(\slashed p+\slashed k) + \lambda} (\gamma^\mu) \frac{1}{\textrm{i}\slashed p + \lambda} \,.
\end{align}
\Cref{eq:SimplifiedSTI-Vertex} is unique up to general transverse functions. For a QED version of this vertex, see \cite{Delbourgo:1977jc,Jia:2017niz}.

\section{Self-energy calculations}
\label{sec:selfenergy-calculations}

In Appendix we collect the results of the momentum integrals for the quark self energy contributions. The momentum integral  \labelcref{eq:DefofI} has been computed in \cite{Horak:2022aza} within dimensional regularisation, and we arrive at
\begin{subequations}
\begin{align}\nonumber
     I_1(p,\lambda_A,\lambda_q)  =&\, \int_{q} \gamma^\mu  \frac{\Pi_{\mu \nu}(q)(-\textrm{i} (\slashed p + \slashed q) + \lambda_q) }{(q^2  + \lambda_A^2)( ( p+q)^2 + \lambda_q^2)} \gamma^\nu \\[2ex]
        &\hspace{-1cm} = \lambda_q I_1^{(s)}(p,\lambda_A,\lambda_q) + \textrm{i} \slashed p  \ I_1^{(d)}(p,\lambda_A,\lambda_q)\,, 
\label{eq:innerintegral1}
\end{align}
with the scalar and Dirac parts of the integrals 
\begin{align}\nonumber 
        I_1^{(s)}(p,\lambda_A,\lambda_q) =& \,\frac{3 \lambda_q}{(4\pi)^2}\bigg(\frac{1}{\epsilon} + \text{log}\,\mu_r^2 - f_0 \bigg)\,,\\[2ex]\nonumber
        I_1^{(d)}(p,\lambda_A,\lambda_q) =&\, \frac{-1}{(4\pi)^2} \Biggl[\bigg(\frac{1}{\epsilon} + \text{log}\,\mu_r^2 \bigg) \sum_{i=0}^3 \frac{\alpha^1_i + \beta^1_i}{i+1} \\[1ex]
                                         &\hspace{1.3cm}- \sum_{i=0}^3 \Bigl(\alpha_1^i f_i + \beta_1^i g_i\Bigr) \Biggr]\,.
\label{eq:scalarDirac1}\end{align}
\end{subequations}
The functions $f_i$ and $g_i$ are given by the weighted Feynman integrals over the logarithms of the cut function $\Delta(\lambda_1,\lambda_2,x,p^2)=  x\left(1-x\right)p^2 + x\left(\lambda_2^2-\lambda_1^2\right) + \lambda_1^2 \,,$
\begin{align}\nonumber
    f_i =&\,\int_0^1 dx \, x^i \,  \log\bigg(\Delta(\lambda_A,\lambda_q,x,p^2)\bigg), \\[1ex]
    g_i =&\,\int_0^1 dx\, x^i \, \log\bigg(\Delta(0,\lambda_q,x,p^2)\bigg)
\end{align}
and the respective prefactors $a_1^i$ and $b^i_1$ read in the Landau gauge
\begin{align}\nonumber
    \alpha_1^0= &\, -2\,, & \alpha_1^1=& \,4  - \frac{p^2 + \lambda_q^2}{\lambda_A^2}\,, & \alpha_1^2=&\,\frac{3p^2}{\lambda_A^2}\,, \\[2ex]
    \beta_1^0=  &\, 0\,, & \beta_1^1=&\,\frac{p^2 + \lambda_q^2}{\lambda_A^2}\,, & \beta_1^2 =&\, \frac{-3p^2}{\lambda_A^2}\,.
\end{align}
The Dirac part of the momentum integral is finite, which reflects the fact that the one-loop anomalous dimension vanishes in the Landau gauge, 
\begin{align}
    \sum_{i=0}^3 \frac{\alpha^1_i + \beta^1_i}{i+1} = 0 \, .
\end{align}
Note that the loop integral is not finite beyond the one-loop approximation, and the respective divergence is carried by the spectral integral. This divergence and the one of the scalar part are cancelled by the BPHZ-subtraction in \cref{eq:sigmaren}, using the spectral BPHZ-scheme introduced in \cite{Horak:2020eng}. Within this scheme, the renormalised self energy follows as 
\begin{align}\nonumber
    \Sigma_q^{\textrm{\tiny{ren}}}(p) = &\, g_s^2\, C_f Z_1  \!\!\int\limits_{\lambda_q,\lambda_A}\!\!\!\!  \lambda_A\,\bar{\rho}_A(\lambda_A) \rho_q(\lambda_q) \\[1ex]\nonumber
   & \hspace{-1cm}\times  \Biggl[ \textrm{i} \slashed p  \Bigl[I_1^{(s)}(p,\lambda_A,\lambda_q) -I_1^{(s)}(\mu_r,\lambda_A,\lambda_q) \Bigr] \\[1ex]
    &  \hspace{-.4cm}+ \lambda_q \Bigr[ I_1^{(d)}(p,\lambda_A,\lambda_q) -I_1^{(d)}(\mu_r,\lambda_A,\lambda_q)\Bigr]  \Biggr] \,,
\label{eq:renormalisedselfenergy}
\end{align}
with
\begin{align}\nonumber
    I_1^{(s)}(p,\lambda_A,\lambda_q)  =&\,-\frac{3}{(4\pi)^2} f_0  \,,\\[1ex]
    I_1^{(d)}(p,\lambda_A,\lambda_q) =&\,\frac{1}{(4\pi)^2}  \sum_{i=0}^3 (\alpha_1^i f_i + \beta_1^i g_i) \,.
    \label{eq:final1}
\end{align}
The inverse Wick rotation is done analytically and the imaginary parts of the self energy read
\begin{align}\label{eq:finalIm}\nonumber
   \textrm{Im}& \, I_1^{(s)}(-\textrm{i}\omega^+,\lambda_A,\lambda_q)  = \frac{\textrm{sign}(w)3\lambda_q}{16 \pi} \theta(\omega-\lambda_A -\lambda_q) \frac{\xi_\textrm{cut}}{\omega^2}\,,\\[2ex]
   \textrm{Im}& \, I_1^{(d)}(-\textrm{i}\omega^+,\lambda_A,\lambda_q) = \frac{\textrm{sign}(w)}{16 \pi} \times  \\[2ex]\nonumber
    &\bigg[\theta(\omega-\lambda_q) \frac{\left( \lambda_q^2 - \omega^2 \right)^3}{2\lambda_A^2\omega^4} + 
  \theta(\omega-\lambda_A -\lambda_q) \frac{\xi_\textrm{cut} l_1 }{2\lambda_A^2\omega^4}\bigg]\,,  \\[2ex]\nonumber
\end{align}
with
\begin{align}\nonumber
    &\xi_{\textrm{cut}} = \sqrt{(\omega^2 - (\lambda_q + \lambda_A)^2) (\omega^2 - (\lambda_q -\lambda_A)^2) } \,,\\[2ex]\nonumber
    &l_1 = \bigg(\lambda_q^4 - 2  \lambda_A^4 + \lambda_A^2 \omega^2 + \omega^4 + \lambda_q^2  (\lambda_A^2 - 2 \omega^2)\bigg)\,.
\end{align}
%

\section{Numerical implementation}
\label{sec:numerics}

In this Appendix, we discuss the numerical solution of the spectral gap equation. All numerical calculations have been carried out in Julia 1.10.4 \cite{julia}. Numerical integrations are based on q-adaptive quadrature  (QuadGK.jl) and cubature rules (HCubature.jl) for one and two-dimensional integrals, respectively. The computation of the self-energy simplifies by using its analytic properties and causal structure. In  particular, the Kramers-Kronig relations entail that the real part of the self-energy is up to a constant given by a principal value integral over its imaginary part and vice versa. Hence, we only compute the finite imaginary parts of the self-energy, see \labelcref{eq:renormalisedselfenergy,eq:finalIm}, and use  the subtracted Kramers-Kronig relations, 
\begin{align}\nonumber
    \text{Re }  \Sigma_{d/s}(\omega) - \text{Re } \Sigma_{d/s}(\textrm{i} \mu_r)& \\[1ex]
         &\hspace{-3cm}=\text{ PV }\int_0^\infty dt \frac{2 t}{\pi} \frac{(\mu_r^2 + \omega^2)\text{Im } \Sigma_{d/s}(t)}{(t^2 + \mu_r^2)(t^2 - \omega^2)}\,. 
\label{eq:KuK}
\end{align}
In \labelcref{eq:KuK} we imposed the Euclidean renormalisation condition at $\textrm{i} \mu_r$. For the numerical integration, we use a fixed but non-linear grid in the variable $\log_{10}[(w-m_\text{pole})/\text{GeV}]$ in the range $[-3,2]$. It is constructed by choosing a grid-spacing of polynomial form with a cubic and linear term to achieve a higher resolution around the peaked structure of the self energies induced by the peak of the effective coupling. The non-linear transformation of the grid is easily inverted for applying standard cubic interpolation methods on a linearly spaced grid, and we use 400 points in frequency direction for the imaginary part of the self energy. The numerical implementation of the Kramers-Kronig relations require an extrapolation of the imaginary parts beyond the grid points. A discontinuity of the integrand at the boundaries of the integration domain leads to a logarithmic divergence of the Principal Value integral at the same scale. For the extrapolation, we use a power-law form. To solve the gap equation, we have employed a fix-point iteration procedure. For an initial enhancement factors $ \eta_{0}= 0.601$ with a small size of chiral symmetry breaking we initialise the iteration with a classical spectral function
\begin{align}
    \rho_q(\lambda>0) = \pi \delta(\lambda - m_q)\,,
    \label{eq:ClassInitialSol}
\end{align}
and iterate until convergence. For successively larger values of $ \eta_{\textrm{\tiny{const}}} $, we use the converged solution obtained for the next smaller value of $ \eta_{\textrm{\tiny{const}}} $ as the input. This leads to a rapid convergence of the iteration scheme and starts the iteration in the basin of attraction of the chirally broken solution. Using \labelcref{eq:ClassInitialSol} in the presence of a large dynamical chiral symmetry breaking may destabilise the iteration, and we may not start in the basin of attraction of the chirally broken solution. 

The onset of the scattering continuum starts exactly at the pole mass. Therefore, a mismatch of pole mass and onset is always present for iteration steps which shift the latter to higher values. This mismatch is regularised by introducing a small shift of the imaginary parts of the self energies at spectral values above the new pole mass. This shift is implemented by cutting off the remnants of the pole at $m_\text{pole} + \gamma$, where the width $\gamma$ is given by the absolute value of the imaginary part of the mass function on the pole. The final spectral function is then approached in the limit $\gamma \to 0$.

\bibliography{spectral_quark_causalvertex.bib}
\end{document}